\documentclass[reprint,preprintnumbers,amsmath,amssymb,aps,pra,floatfix]{revtex4-1}
\usepackage{amsmath}
\usepackage{amssymb}
\usepackage{mathtools}
\usepackage{bbold}
\usepackage[dvipsnames,usenames]{xcolor}
\usepackage[qm,braket]{qcircuit}
\usepackage[margin=2cm]{geometry}
\usepackage{tikz}
\usetikzlibrary{automata,arrows,positioning,calc}
\usepackage[utf8]{inputenc}
\usepackage{array}
\usepackage{lipsum}
\usepackage{empheq}
\usepackage{hyperref}
\hypersetup{
    colorlinks=true,
    linkcolor=blue,
    citecolor=blue,
    }
\usepackage[nameinlink]{cleveref}
\usepackage[all]{hypcap}
\usepackage[normalem]{ulem}

\newcommand{\bb}[1]{\mathbb{#1}} 

\newcommand{\UW}{{Department of Physics, University of Washington, Seattle, WA 98195, USA}}
\newcommand{\IQUS}{{InQubator for Quantum Simulation (IQuS), Department of Physics, University of Washington, Seattle, WA 98195, USA}}
\newcommand{\UNITN}{{Dipartimento di Fisica, University of Trento, via Sommarive 14, I–38123, Povo, Trento, Italy}}
\newcommand{\TIFPA}{INFN-TIFPA Trento Institute of Fundamental Physics and Applications,  Trento, Italy}
\newcommand{\UT}{{Department of Computer Science, University of Toronto, Toronto, ON M5S 2E4, Canada}}
\newcommand{\PNNL}{{Pacific Northwest National Laboratory, Richland, WA 99354, USA}}

\begin{document}

\title{Quantum Error Correction with Gauge Symmetries}
%
%
%
\author{Abhishek Rajput}
\email{rajpua@uw.edu}
\affiliation{\UW}

\author{Alessandro~Roggero}
\email{a.roggero@unitn.it}
\affiliation{\IQUS}
\affiliation{\UNITN}
\affiliation{\TIFPA}

\author{Nathan~Wiebe}
\email{nawiebe@cs.toronto.edu}
\affiliation{\UT}
\affiliation{\PNNL}
\affiliation{\UW}

%
%
%
\preprint{IQuS@UW-21-014}
%

\begin{abstract}


Quantum simulations of Lattice Gauge Theories (LGTs) are often formulated on an enlarged Hilbert space containing both physical and unphysical sectors in order to retain a local Hamiltonian.  We provide simple fault-tolerant procedures that exploit such redundancy by combining a phase flip error correction code with the Gauss' law constraint to correct one-qubit errors for a $\mathbb{Z}_2$ or truncated U(1) LGT in 1+1 and 2+1 dimensions with a link flux cutoff of $1$. Unlike existing work on detecting violations of Gauss' law, our circuits are fault tolerant and the overall error correction scheme outperforms a na\"{i}ve application of the $[5,1,3]$ code. The constructions outlined can be extended to LGT systems with larger cutoffs and may be of use in understanding how to hybridize error correction and quantum simulation for LGTs in higher space-time dimensions and with different symmetry groups.

\end{abstract}
\maketitle

\section{Introduction}

As gauge theories lie at the heart of the framework governing the interactions and forces described by the Standard Model, considerable effort has been expended in numerical simulations of these theories for computing physical quantities. Lattice gauge theories (LGTs) have been among the most fruitful formulations of non-perturbative approaches amenable for implementation on classical computers~\cite{Wilson1974,Kogut1975,Kogut1979,Durr2008,Kronfeld2012}. In regimes where classical simulations are plagued by an exponential scaling of the computational cost, digital quantum computers have emerged as a promising platform for the efficient simulation of LGTs. Notable examples are real-time dynamics or the study of systems at finite density~\cite{Zohar2015,Dalmonte2016,Banuls2020,klco2021standard}.

Despite considerable advances on this front, digital quantum simulation of LGTs can time-evolve a given fiducial state into unphysical sectors due to noise and the approximation error associated with the simulation protocol used. A popular approach to mitigate this problem, especially useful for ground-state calculations but shown to be useful also out-of-equilibrium~\cite{Halimeh2020}, is to enforce Gauss' Law by adding an energy penalty to the Hamiltonian (see e.g~\cite{Dalmonte2016,Hauke2013,Zohar2015,vandamme2020gaugesymmetry,Halimeh2020} and also~\cite{rajput2021hybridized} for techniques to reduce the gate cost of adding penalty terms). Another recent proposal for error mitigation in gauge theories uses random gauge transformations to suppress the component of the quantum state in the unphysical Hilbert space~\cite{lamm2020suppressing,Tran2021}. The recent approach proposed in~\cite{StrykerOracle_2019}, and its generalization to non-Abelian theories~\cite{Raychowdhury2020}, uses instead a quantum oracle to detect the presence of gauge violating errors by performing explicit Gauss' Law checks and flagging an ancilla qubit. These techniques are suitable for error detection but in general do not possess error correction capabilities and are not fault-tolerant. 

With the long-time goal of performing quantum simulation of LGT on fault-tolerant protocols, an intriguing possibility is to tailor general purpose error correction schemes to best exploit the structural properties of these theories in order to reduce the resource requirements for early explorations (see e.g.~\cite{klco2021hierarchical} for a recent attempt in this direction using the surface code). The physical intuition behind the approach followed in this work is that error correcting codes can be seen as artificial gauge theories where the logical Hilbert space is determined by states that satisfy a suitable local symmetry. When simulating LGTs which themselves need to satisfy a physical local symmetry, it might then be advantageous to exploit this natural redundancy to reduce the cost of the full error correction encoding.

We develop in this work fault-tolerant algorithms for error correction suitable for $\mathbb{Z}_2$ or truncated U(1) lattice gauge theories in 1+1, and 2+1 spacetime dimensions with a cutoff of $1$ on the links. These algorithms combine the physicality constraint provided by Gauss' Law with bit and phase flip error correction codes~\cite{Shor1995} to detect and correct errors stemming from device noise that occur on a site or its adjacent links. Error correction for a 1D lattice with $2N$ links and $2N$ staggered fermionic sites with periodic boundary conditions is accomplished by tessellating these encodings across the whole lattice so that errors occurring on a particular site and its adjacent link can be detected by a Gauss' Law check on the next set of sites and links. The extension to 2D follows a similar idea using instead plaquettes and links coming out of them as the fundamental blocks of the partition.

\begin{table}
\centering 
    \begin{tabular}{ | >{\centering\arraybackslash} m{1.8cm} | >{\centering\arraybackslash} m{2cm} | >{\centering\arraybackslash} m{1.5cm} | }
        \hline
        \textbf{1+1} & Pure Gauge & Dynamical \\ 
        \hline
        [5,1,3] & $10N$ & $20N$ \\ 
        \hline
        Gauss's Law (Doubling) & $9N$ & $15N$ \\
        \hline
        Gauss's Law & $6N$ & $12N$ \\
        \hline
        \textbf{2+1} & & \\
        \hline
        [5,1,3] & $40 N_x N_y$ & $60 N_x N_y$ \\
        \hline
        Gauss's Law & $36 N_x N_y$ & $48 N_x N_y$ \\
        \hline
    \end{tabular}
        \caption{Number of qubits required (excluding ancillas) for performing fault-tolerant error-correction with different encodings on a $\bb{Z}_2$ or truncated U(1) 1+1 dimensional LGT system with $2N$ links, $2N$ staggered fermions and a 2+1 dimensional LGT with $8 N_x N_y$ links and $4 N_x N_y$ sites, both with a flux cutoff of $1$. The second row gives the cost for an encoding where Gauss's law is exploited to give a bit-flip encoding via an extra even-numbered link qubit in 1+1 D. The last row of the 1+1 and 2+1 D cases gives the cost when only the redundancies from Gauss's Law are used at the logical level to perform error correction. The case with non-dynamical fermions requires the same resources as the pure gauge case and is omitted from the table.}
    \label{table:EncodingCost}
\end{table} 

\begin{figure}
\includegraphics[width=0.5\textwidth]{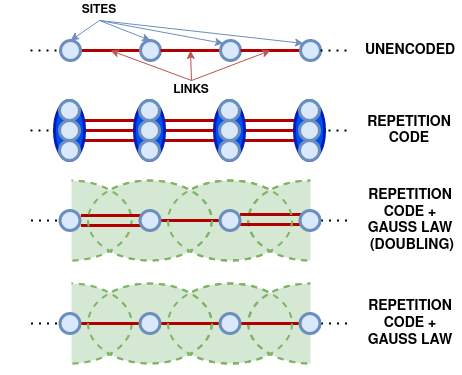}
\caption{Schematic illustration of the differences between two error correcting schemes for a simple one-dimensional LGT: a traditional bit-flip encoding scheme and the two schemes proposed here exploiting the Gauss' Law gauge symmetry.  \label{fig:ecpic}}
\end{figure}

In the electric basis used here, a gauge violation is caused by bit-flip errors which can be corrected using a standard encoding based on the repetition code using $12N$ qubits: $3$ for each site and $3$ for each link (see the schematic illustration in \autoref{fig:ecpic}). The error correction procedure proposed here instead requires no bit-flip repetition code for the fermionic sites and a compressed encoding requiring only half of the qubits ($3N$ in total) for the link variables. Full fault-tolerance can then be achieved by concatenation with a standard phase-flip code (similar to what is done in the 9 qubit Shor code~\cite{Shor1995}) and using fault tolerant gadget design (e.g. using flag qubits~\cite{Chao_2018}). For the simpler case when the fermionic sites are non-dynamical, i.e. they can be represented by classical bits, the scheme proposed here requires $9N$ qubits for unit flux cutoffs with O(1) ancillas. The full construction depicted in \autoref{fig:ecpic} involving dynamical fermions requires $15N$ qubits with O(1) ancillas. These are to be contrasted with the $10N$ and $20N$ qubits required for the two cases respectively using the [5,1,3] code for each site and link~\cite{DiVincenzo1996,Laflamme1996}. Similarly, our construction for the 2D case requires for a lattice with $8N_xN_y$ links and $4N_xN_y$ sites, a total of $48N_xN_y$ qubits for the dynamical fermion case and $36N_xN_y$ qubits in pure gauge or with only static charges. For comparison, a $[5,1,3]$ code would require $60N_xN_y$ and $40N_xN_y$ qubits respectively. It is foreseeable that more space efficient codes could be constructed, with the present work developing a first attempt to exploit the intrinsic redundancy present in gauge theories to reduce the cost of the error correction procedures required for their simulation at large scale.

The structure of the paper is as follows: \Cref{sec:abelian} introduces the structure of U(1) and $\mathbb{Z}_n$ Abelian LGTs. \Cref{sec:errorcorrection} surveys the basics of error correction codes and develops the formalism for integrating Gauss' law with a repetition code. \Cref{sec:applications} then presents applications of these ideas for pure gauge theory first before treating the cases of error correction for a $Z_2$ 1+1 dimensional LGT with both non-dynamical and dynamical fermions. We discuss an extension of these ideas to the more complex case of 2 spatial dimensions in \Cref{sec:2dlgt}. \Cref{sec:discussion} summarizes the results and discusses future work for extending these results to different symmetry groups, and higher spacetime dimensions and flux cutoffs. 

\section{Structure of U(1) and \texorpdfstring{$\mathbb{Z}_n$}{} Abelian Lattice Gauge Theories}
\label{sec:abelian}

We follow the basic outline given in \cite{StrykerOracle_2019} and review the structure of abelian lattice gauge theories, specifically for the general gauge groups $G=\bb{Z}_n$ and $G= $ U(1) which contain those considered throughout this work. There are several physical models for which the gauge symmetries discussed here are important such as the Schwinger Model, or QED in 1+1 dimensions on a lattice \cite{Schwinger1962,COLEMAN1975}. It is the one of the simplest concrete examples of an Abelian LGT and serves as a convenient setting for the analysis and application of Gauss' Law symmetries to error correction. This model has been extensively used as an important stepping stone in simulations of lattice field theories using both tensor networks and quantum devices \cite{Ba_uls_2013, Pichler2016, Hauke2013, Martinez_2016, Klco2018}. 

We discretize space on a cubic lattice $L$ with sites labeled by $s$ and links labeled by $l$.  We assume that the lattice consists of $N$ sites for even $N\ge 0$ and that a staggered fermion representation is used wherein every second site is positronic.  Each link $l$ is associated an independent separable Hilbert space $\mathcal{H}_l$ with the same orthonormal basis:
\begin{equation}
   \bra{\epsilon'} \epsilon \rangle = \delta_{m',m},  \hspace{3em} \hat{\mathbb{1}} = \sum_{\epsilon} \ket{\epsilon}\bra{\epsilon} 
\end{equation}
with
\[
    \epsilon',\epsilon \in 
\begin{cases}
    \bb{Z}_n, & \text{if } G = \bb{Z}_n \\
    \bb{Z},              & \text{if } G = \text{U}(1).
\end{cases}
\]

The Hamiltonian for this LGT is a function of the link operators $\hat{U}_l$ and their conjugate electric fields $\hat{E}_l$ defined explicitly in this basis by 
\begin{equation}
    \hat{U}_l = \sum_{\epsilon_l} \ket{\epsilon_l+1}\bra{\epsilon_l}, \hspace{2em} \hat{E}_l = \sum_{\epsilon_l} \epsilon_l \ket{\epsilon_l}\bra{\epsilon_l}. \label{eq:linkops}
\end{equation}
From this expression we can see that $\hat{U}_l$ acts as a raising operator and its adjoint as a lowering operator on the Hilbert space $\mathcal{H}_l$ of the link. Operators defined on different link Hilbert spaces commute while the same-link commutation relations are given by
\begin{align}
    [\hat{E}_l, \hat{U}_l] &= \hat{U}_l, & G &= \mathrm{U}(1) \label{eq:samecommU1}\\
    \hat{Q}_l \hat{U}_l \hat{Q}_l^{\dag} &= \hat{U}_l e^{2 \pi i/n} & G &= \bb{Z}_n \label{eq:samecommZn}
\end{align}

where $$\hat{Q}_l \coloneqq e^{2 \pi i \hat{E}_l/n} = \sum_{\epsilon_l =0}^{N-1} e^{2 \pi i \epsilon_l/n} \ket{\epsilon_l} \bra{\epsilon_l}.$$ The form of the commutation relation for $\bb{Z}_n$ is due to the fact that the electric field values are periodic, so the Hamiltonian depends on $\hat{Q}_l$.

When considering fermionic matter fields on the sites, we work in the occupation number basis where number operators $n_\sigma$ are diagonal with eigenvalues $\{0,1\}$. Here, $\sigma$ is a collective index denoting the relevant species or indices (like flavor or spinor) involved. The global state of the entire lattice is spanned by a basis given by a specification of electric fields on the links of the lattice and occupation numbers on the sites. Due to the locality of the symmetry, we will typically consider a particular site on a lattice and those links attached to it and denote the corresponding basis states by $$\ket{\mathbf{E},\rho} \rightarrow \otimes_{i=1}^D \ket{E_i(s)} \otimes_{i=1}^D \ket{E_i(s-\hat{e}_i)} \otimes_{\sigma} \ket{n_{\sigma}},$$ where $D$ is the spatial dimension of the lattice and $\rho$ is a discretized charge density defined by $$
\hat{\rho}(s) = \sum_{\sigma} e_{\sigma} \hat{n}_{\sigma}(s),$$ and $e_{\sigma} = \pm 1$. Due to gauge invariance, states in the physical Hilbert space satisfy a local Gauss' Law which relates the state of a site with the state of the links emanating from it. Gauge-invariant states are in the kernel of the operator 
\begin{align}
\begin{split}
    \hat{G}_s & \coloneqq (\nabla \cdot \hat{E})(s) - \hat{\rho}(s)  \\
    &\coloneqq \sum_{i=1}^D (\hat{E}_i(s) - \hat{E}_i(s-\hat{e}_i)) - \sum_{\sigma} e_{\sigma} \hat{n}_{\sigma}(s)\;, \label{eq:gausslaw}
\end{split}
 \end{align}
where the second line is obtained by discretization of the gradient operator on the lattice. 

When dealing with a U(1) gauge group, it is necessary to truncate the link electric field values to enable digital quantum simulation with a finite number of qubits. This can be accomplished by "wrapping" the electric field at a cutoff $\Lambda$: 
\begin{align}
&\hat{E}_l = \sum_{\epsilon_l = -\Lambda}^{\Lambda -1} \epsilon_l \ket{\epsilon_l}\bra{\epsilon_l} \label{eq:elecwrap}\\
&\hat{U}_l\ket{\Lambda -1} = \ket{-\Lambda} \label{eq:raisingwrap}\\
&\hat{U}_l^{\dag}\ket{-\Lambda} = \ket{\Lambda-1}. \label{eq:loweringwrap}
\end{align}

This choice of discretization results in a modification of the commutation relations as follows:
\begin{align}
    [\hat{E}_l, \hat{U}_l] &= \hat{U}_l - 2\Lambda\ket{-\Lambda} \bra{\Lambda -1} \\
    [\hat{E}_l, \hat{U}_l^{\dag}] &= -\hat{U}_l^{\dag} + 2\Lambda\ket{\Lambda-1} \bra{-\Lambda}.
\end{align}

Note that with our choice of the lower and upper bound, the link Hilbert spaces are even-dimensional and can therefore be mapped onto a $\lceil\log(2\Lambda)\rceil$-qubit Hilbert space. In this work, we will restrict the discussion to $\Lambda=1$ and comment on the prospects of generalizing our constructions to arbitrary cutoffs in~\Cref{sec:discussion}.

\section{Error Correction with Gauss' Law}

\label{sec:errorcorrection}

We now briefly discuss the bit and phase-flip error correction codes used throughout the paper and show how to integrate them with Gauss Law to reduce the number of qubits required. The overarching idea is to perform an encoding of the link physical qubits into logical states and use the local gauge symmetry to implement a more space efficient error correction code under an error model with arbitrary single-qubit errors. 

\subsection{Preliminaries on the repetition code}

We outline in this section the basics of the bit and phase flip error correction codes following the treatment given in \cite{MikeIke}.
First consider a noisy classical communications channel through which we wish to send a bit between two locations and suppose its behavior is such that it flips the bit with probability $p$. To protect the bit against the effects of noise, we can employ what is known as a "repetition code". This involves replacing the bit with three copies of itself, i.e. $0 \rightarrow 000$ and $1 \rightarrow 111$. These new bit strings are denoted as the ``logical 0" and ``logical 1" and we send these through the channel. The receiver then attempts to decode what the original bit was. If the output is $010$ for instance, then provided the probability $p$ of error is not high and the noise acts independently on each bit, it is likely the second bit was flipped and that the original bit was $0$. This is known as majority voting, since the intended original message is determined by whatever bit value appears more in the output. This can obviously fail if more than one bit was flipped. It can be easily determined that with this encoding scheme, the transmission becomes more reliable if $p < 1/2$. 

\begin{figure}[b]
    \[
            \Qcircuit @C=2em @R=2em {
            \lstick{\ket{\psi}} & \ctrl{1} & \ctrl{2} & \qw \\
            \lstick{\ket{0}} & \targ & \qw & \qw \\
            \lstick{\ket{0}} & \qw & \targ & \qw 
            }
    \]
\caption{Circuit for creating the encoded logical state $\ket{\psi}_L = a\ket{000} + b\ket{111}$ for the bit-flip code.} \label{fig:bitfliplogstate} 
\end{figure}
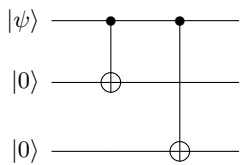

Now consider a noisy quantum channel that applies a bit-flip, or $X$ gate, to a state $\ket{\psi}$ sent through it with probability $p$. We write $\ket{\psi}$ in terms of the computational basis as $\ket{\psi} = a\ket{0} + b\ket{1}$ and encode it in three qubits as $a \ket{000} + b\ket{111}$. In other words, we have a mapping between the "physical" qubits $\ket{0}$ and $\ket{1}$ to the logical qubits $\ket{0}_L = \ket{000}$ and $\ket{1}_L = \ket{111}$ respectively, where the subscript $L$ denotes a logical state. Such an encoding can be accomplished by the circuit in \autoref{fig:bitfliplogstate}.

Each qubit in the encoded state is passed through a separate bit-flip channel. If a bit-flip occurs on at most one qubit, we can measure the parities of the qubits by performing projective measurements of the operators $Z_1 Z_2$ and $Z_2 Z_3$, where the tensor product is implied. This process is known as making ``syndrome measurements" for the error syndromes $Z_1 Z_2$ and $Z_2 Z_3$. These operators have eigenvalues of $\pm 1$. $Z_1 Z_2$ measures the parities of the first two qubits and yields the eigenvalue $-1$ if they differ and $+1$ if they do not. $Z_2 Z_3$ acts in the same way for the second and third qubits. The measurement outcomes of either operator allows us to determine which qubit was flipped. For instance, if eigenvalues of $-1$ are obtained from the measurement of both operators, we know that with high probability the second qubit was flipped. We can then perform error correction by applying an $X$ gate on the $2^{\rm nd}$ qubit to flip it back to its original state. Note that the measurement of these operators gives no information about the amplitudes $a$ and $b$ of the encoded state and therefore do not destroy the state we wish to perform error detection and correction on. 
\Cref{table:bitflipsyndrome} outlines the possible measurement outcomes for the syndromes $Z_1 Z_2$ and $Z_2 Z_3$ and the error correction operations to perform. \autoref{fig:bitflipECcirc} gives the full circuit to correct bit-flip errors. An equivalent circuit used for the projective measurement of the stabilizers $Z_1 Z_2$ and $Z_2 Z_3$ is presented in \autoref{fig:StabMeasEquiv}. We will use this decomposition in the rest of this work.

\begin{table}
\centering 
    \begin{tabular}{ | m{1cm} | m{1cm}| m{2cm} | }
        \hline
        $Z_1 Z_2$ & $Z_2 Z_3$ & Correction \\ 
        \hline
        1 & 1 & III \\ 
        \hline
        1 & -1 & IIX \\ 
        \hline
        -1 & 1 & XII \\ 
        \hline
        -1 & -1 & IXI \\ 
        \hline
    \end{tabular}
\caption{Syndrome measurements outcomes, and correction operations for bit flip error correction code.}
\label{table:bitflipsyndrome}
\end{table} 

\begin{figure}[b]
    \[
        \Qcircuit @C=0.7em @R=1em {
        & \mbox{Encoding} & & & & & & & \mbox{Recovery}\\
        \lstick{\ket{\psi}} & \ctrl{1} & \ctrl{2} & \gate{Z}  & \qw & \qw & \qw & \gate{X} & \qw & \qw & \qw \\
        \lstick{\ket{0}} & \targ & \qw & \gate{Z} \qwx & \gate{Z} \qw & \qw & \qw & \qw & \gate{X} & \qw & \qw \\
        \lstick{\ket{0}} & \qw & \targ & \qw & \gate{Z} \qwx \qw & \qw & \qw & \qw & \qw & \gate{X} & \qw \\ 
        \lstick{\ket{0}} & \gate{H} & \qw & \ctrl{-2} & \qw & \gate{H} & \meter & \cctrl{-3} & \cctrl{-2} & \cctrlo{-1} \\
        \lstick{\ket{0}} & \gate{H} & \qw & \qw & \ctrl{-2} & \gate{H} & \meter & \cctrlo{-1} & \cctrl{-1} \gategroup{2}{1}{4}{3}{.7em}{--} & \cctrl{-1} \gategroup{2}{8}{6}{11}{.7em}{--} & &
        }
    \]
    \caption{Full circuit for correcting a bit flip errors on a general state $\ket{\psi}$.}
    \label{fig:bitflipECcirc}
\end{figure}
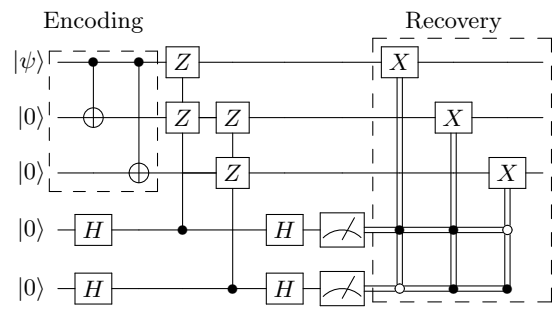

\begin{figure}[t]
\vspace{-0.5cm}
\[
    \Qcircuit @C=1em @R=1.5em {
    \lstick{\ket{\psi}} & \ctrl{3} & \qw & \qw & \qw & \qw \\
    \lstick{\ket{0}} & \qw & \ctrl{2} & \ctrl{3} & \qw & \qw \\
    \lstick{\ket{0}} & \qw & \qw & \qw & \ctrl{2} & \qw \\
    \lstick{\ket{0}} & \targ & \targ & \qw  & \qw & \qw \\
    \lstick{\ket{0}} & \qw & \qw & \targ & \targ & \qw 
    }
\]
\label{fig:StabMeasEquiv}
\caption{Equivalent circuit for the measurement of the stabilizers $Z_1 Z_2$ and $Z_2 Z_3$ for the bit-flip error correction code. This can be obtained from the identity $HZH = X$ and the fact that controlled-Z gates are equivalent to a controlled-Z gate with the control and target flipped.}
\end{figure}
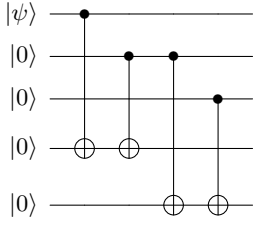

Now suppose we have a noisy quantum channel that applies a phase flip (i.e a $Z$ gate) with probability $p$ to a qubit in the state $\ket{\psi} = a\ket{0} + b\ket{1}$. Unlike the bit-flip encoding, there is no classical analogue of applying a ``phase" to a bit. However, we can convert this channel to a bit flip channel by working in the $\ket{+} = (\ket{0} + \ket{1})/\sqrt{2}$ and $\ket{-} = (\ket{0}-\ket{1})/\sqrt{2}$ basis. With respect to this basis, the $Z$ operator takes $\ket{+}$ to $\ket{-}$ and therefore acts as a bit flip. 
We can then apply the same logic for error correction in the bit-flip case to the present case by switching from the computational basis to the $\ket{+}$ and $\ket{-}$ basis via the Hadamard gate. In our present work, we will find it convenient to use the logical codeword basis $\ket{0}_L = (\ket{+++} + \ket{---})/\sqrt{2}$ and $\ket{1}_L = (\ket{+++} - \ket{---})/\sqrt{2}$. Then to detect errors, we can perform projective measurements of the stabilizers $X_1 X_2$ and $X_2 X_3$ to determine the parity of the bits. Based on the measurement outcomes, we can apply $Z$ gates to correct the errors accordingly. The encoding circuit and the phase-flip error correction procedure are depicted in \Cref{fig:phasefliplogstate} and \Cref{fig:phaseflipECcirc} respectively.
\begin{figure}[b]
    \[
            \Qcircuit @C=2em @R=2em {
            \lstick{\ket{\psi}} & \gate{H} & \ctrl{1} & \ctrl{2} & \gate{H} & \qw \\
            \lstick{\ket{0}} & \qw & \targ & \qw & \gate{H} & \qw \\
            \lstick{\ket{0}} & \qw & \qw & \targ & \gate{H} & \qw
            }
    \]
\caption{Circuit for creating the encoded logical state $\ket{\psi}_L = a\ket{0}_L + b\ket{1}_L$ for the phase-flip code, where $\ket{0}_L = (\ket{+++} + \ket{---})/\sqrt{2}$ and $\ket{1}_L = (\ket{+++} - \ket{---})/\sqrt{2}$.} \label{fig:phasefliplogstate} 
\end{figure}
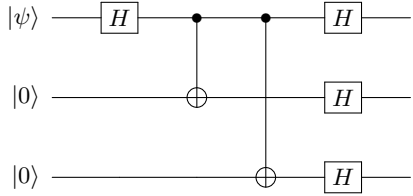 
\begin{figure}
    \[
        \Qcircuit @C=0.7em @R=1em {
        & & \mbox{Encoding} & & & & & & & \mbox{Recovery}\\
        \lstick{\ket{\psi}} & \gate{H} & \ctrl{1} & \ctrl{2} & \gate{H} & \gate{X}  & \qw & \qw & \push{\rule{0em}{1em}} \qw & \gate{Z} & \qw & \qw & \qw \\
        \lstick{\ket{0}} & \qw & \targ & \qw & \gate{H} & \gate{X} \qwx & \gate{X} \qw & \qw & \qw & \qw & \gate{Z} & \qw & \qw \\
        \lstick{\ket{0}} & \qw & \qw & \targ & \gate{H} & \qw & \gate{X} \qwx \qw & \qw & \qw & \qw & \qw & \gate{Z} & \qw \\ 
        \lstick{\ket{0}} & \qw & \qw & \qw & \gate{H} & \ctrl{-2} & \qw & \gate{H} & \meter & \cctrl{-3} & \cctrl{-2} & \cctrlo{-1} \\
        \lstick{\ket{0}} & \qw & \qw & \qw & \gate{H} & \qw & \ctrl{-2} & \gate{H} & \meter & \cctrlo{-1} & \cctrl{-1} \gategroup{2}{1}{4}{5}{.7em}{--} & \cctrl{-1}  \gategroup{2}{9}{6}{13}{.7em}{--} &
        }
    \]
    \caption{Full circuit for correcting phase flip errors on a general state $\ket{\psi}$.} 
    \label{fig:phaseflipECcirc}
\end{figure}
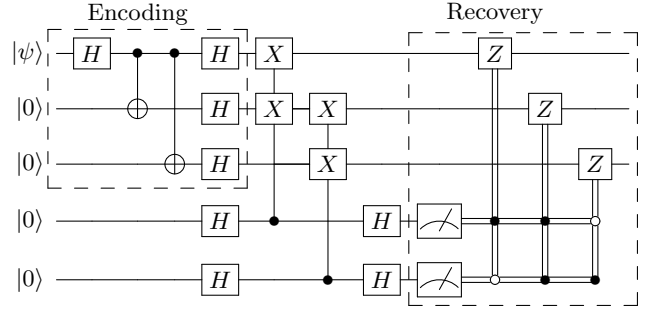

The constructions given in~\cite{Chao_2018} allow us to ensure the fault tolerance of the encoding, error detection, and recovery operations in either code. The underlying technique involves introducing an extra ``flag" qubit prepared in the $\ket{+}$ state and performing CNOT operations from it on the syndrome ancilla qubit at key points in the circuit (see Figure 3(b) in~\cite{Chao_2018}). This flag qubit is then measured in the $X$ basis and a result of $\ket{-}$ indicates an error of weight two or more on the data qubits. Additional flag qubits can be added between each gate in the stabilizer measurement to ensure a localization of errors. Similar constructions apply in creating fault-tolerant versions of other important subroutines like logical state-preparation and stabilizer measurements (see Appendix A in~\cite{Chao_2018}).

\subsection{Integrating the repetition code with Gauss' Law}
\label{subsec:gauss_construction}

We are now in the position to discuss how the expression of Gauss' Law in Eq.~\eqref{eq:gausslaw} can remove the need to use an explicit bit flip code in fault tolerant LGT simulations. To simplify the discussion we will henceforth consider the one dimensional case ($D=1$) with staggered fermionic sites, where only one fermionic flavor with charge $e_s$ is present on any given site $s$. The Gauss' Law operator at site $s$ between links $l$ and $l+1$ simplifies then to
\begin{equation}
\hat{G}_s = \hat{E}_{l+1} - \hat{E}_l - e_s\hat{n}(s)\;,
\end{equation}
and in the simpler pure gauge case with no matter to
\begin{equation}
\label{eq:glaw_pureg}
\hat{G}_s = \hat{E}_{l+1} - \hat{E}_l\;.
\end{equation}
We start by discussing the latter case and consider a general state of the two links in the electric basis
\begin{equation}
\label{eq:two_link_state}
\ket{\Psi_{l,l+1}} = \sum_{\epsilon_l}\sum_{\epsilon_{l+1}} \Psi_{\epsilon_{l},\epsilon_{l+1}} \ket{\epsilon_l}\otimes\ket{\epsilon_{l+1}}\;.
\end{equation}
Following the discussion in Sec.~\ref{sec:abelian}, a physical state needs to be in the kernel of the Gauss' Law operator. Using Eq.~\eqref{eq:glaw_pureg} above we see that $\hat{G}_s$ acts on $\ket{\Psi_{l,l+1}}$ as
\begin{equation}
\label{eq:glaw_pureg_state}
\hat{G}_s \ket{\Psi_{l,l+1}} = \sum_{\epsilon_l}\sum_{\epsilon_{l+1}} \Psi_{\epsilon_{l},\epsilon_{l+1}}(\epsilon_{l+1}-\epsilon_{l}) \ket{\epsilon_l}\otimes\ket{\epsilon_{l+1}}\;,
\end{equation}
and is zero only if the coefficient matrix is diagonal
\begin{equation}
\Psi_{\epsilon_{l},\epsilon_{l+1}}(\epsilon_{l+1}-\epsilon_{l})=0\; \Leftrightarrow\; \Psi_{\epsilon_{l},\epsilon_{l+1}} = \Psi_{\epsilon_{l}} \delta_{\epsilon_{l},\epsilon_{l+1}}\;.
\label{eq:constraint}
\end{equation}
This argument, which can be easily generalized to the case where the state of the links is mixed, shows that gauge invariant states are analogous to a bit-flip repetition code with only two copies: $\ket{0}_L=\ket{0}\otimes\ket{0}$ and $\ket{1}_L=\ket{1}\otimes\ket{1}$. The distance between these codewords is not sufficient to allow for error correction but is sufficient for error detection (see e.g.~\cite{Corcoles2015}). This explains in an intuitive way why oracles like those presented in Ref.~\cite{StrykerOracle_2019} are capable of detecting bit-flip errors without requiring additional qubits for the encoding.

It is now easy to see how the use of Gauss' Law allows for a space reduction in the bit-flip encoding: the standard procedure described above will require three link registers to encode a logical link as
\begin{align}
\ket{\Phi_l}_L &\coloneqq \sum_{\epsilon_l} \Phi_{\epsilon_l} \ket{\epsilon_l}\otimes\ket{\epsilon_l}\otimes\ket{\epsilon_l} \nonumber \\
&=\sum_{\epsilon_l} \Phi_{\epsilon_l} \ket{\epsilon_l}_L. \label{eq:bitflip-logical}
\end{align}
The fact that physical states satisfy Eq.~\eqref{eq:glaw_pureg_state} means we need only two registers per link and can use one of the two registers for the $l+1$ link across a site $s$ when we measure stabilizers and perform error recovery. Since only three registers are involved in this procedure, we will consider a construction with two qubit registers for even links and only one register for the odd links in order to minimize the memory cost. More explicitly, we will use the alternative encoding
\begin{equation}
\ket{\Phi_l}_{GLE}\coloneqq \sum_{\epsilon_l} \Phi_{\epsilon_l} \ket{\epsilon_l}\otimes\ket{\epsilon_l}=\sum_{\epsilon_l} \Phi_{\epsilon_l} \ket{\epsilon_l}_{GLE}\;,
\end{equation}
for the even links in the lattice and
\begin{equation}
\ket{\Phi_l}_{GLO}\coloneqq \sum_{\epsilon_l} \Phi_{\epsilon_l} \ket{\epsilon_l}=\sum_{\epsilon_l} \Phi_{\epsilon_l} \ket{\epsilon_l}_{GLO}\;,
\end{equation}
equivalent to a bare encoding, for the odd links. For physical states that satisfy the Gauss' law constraint in Eq.~\eqref{eq:constraint} we have then
\begin{equation}
\begin{split}
\label{eq:gauss_logical_state_pg}
\ket{\Phi_{l,l+1}}_{GL} &= \sum_{\epsilon_l}\sum_{\epsilon_{l+1}} \Psi_{\epsilon_{l},\epsilon_{l+1}} \ket{\epsilon_l}_{GLE}\otimes\ket{\epsilon_{l+1}}_{GLO}\\
&= \sum_{\epsilon_l} \Psi_{\epsilon_{l}} \ket{\epsilon_l}_{GLE}\otimes\ket{\epsilon_{l}}_{GLO}\\
&= \sum_{\epsilon_l} \Psi_{\epsilon_{l}} \ket{\epsilon_l}\otimes\ket{\epsilon_l}\otimes\ket{\epsilon_l}\;,
\end{split}
\end{equation}
and we can now use the stabilizer measurements and recovery operation on the three qubits as in the bit-flip repetition code. The second line is obtained by ensuring that the logical state $\ket{\Phi_{l,l+1}}_{GL}$ is in the kernel of the logical Gauss Law operator derived from Eq.~\eqref{eq:glaw_pureg}
\begin{equation}
\begin{split}
\hat{G}^{(GL)}_{s} &= \hat{E}^{(GLO)}_{l+1} - \hat{E}^{(GLE)}_l\\
&=\hat{E}_{l+1} - \hat{E}_l\otimes\hat{E}_l\;.
\end{split}
\end{equation}
The expressions above apply directly to sites $s$ between $l$ even and $l+1$ odd, but it is straightforward to generalize them to the site $s+1$ where the left link is odd and the right link is even as follows
\begin{align}
\label{eq:gauss_logical_state_pg_oe}
\ket{\Phi_{l+1,l+2}}_{GL} &= \sum_{\epsilon_{l+1}}\sum_{\epsilon_{l+2}} \Psi_{\epsilon_{l+1},\epsilon_{l+2}}  \ket{\epsilon_{l+1}}_{GLO}\otimes\ket{\epsilon_{l+2}}_{GLE} \nonumber \\
&= \sum_{\epsilon_{l+1}} \Psi_{\epsilon_{l+1}} \ket{\epsilon_{l+1}}_{GLO}\otimes\ket{\epsilon_{l+1}}_{GLE} \nonumber \\
&= \sum_{\epsilon_{l+1}} \Psi_{\epsilon_{l+1}} \ket{\epsilon_{l+1}}\otimes\ket{\epsilon_{l+1}}\otimes\ket{\epsilon_{l+1}}\;.
\end{align}
The gauge invariant logical state is again equivalent to the standard bit-flip encoding shown in Eq.~\eqref{eq:bitflip-logical}. Since Gauss' Law only detects errors in the flux value, which in our representation corresponds to bit-flip errors, each register of qubits in the construction presented above needs to be encoded in a phase-flip code to ensure that all single errors are correctable. With this concatenation, the gauge-invariant logical states $\ket{\Phi_{l,l+1}}_{GL}$ and $\ket{\Phi_{l+1,l+2}}_{GL}$ become equivalent to a full 9-qubit encoding~\cite{Shor1995}. However, we encode two full links using the 9-qubit encoding as opposed to each link separately. This encoding therefore has better memory efficiency than error correcting codes that act on individual qubits, the best one requiring $5$ qubits~\cite{DiVincenzo1996,Laflamme1996}. Note that it is still possible to find codes with even higher memory efficiency when multiple logical qubit are encoded. For instance, $3$ logical qubits can be encoded into $8$ while correcting all single errors (see eg.~\cite{Gottesman1996,Calderbank1997}).

Adding fermions to the system modifies the relation between link states across a site so that they do not necessarily need to be the same. The full state around a site $s$ is a generalization of Eq.~\eqref{eq:two_link_state} and should now be written as
\begin{equation}
\ket{\Psi^s_{l,l+1}} = \sum_{\epsilon_l}\sum_{\epsilon_{l+1}}\sum_{n=0,1} \Psi^n_{\epsilon_{l},\epsilon_{l+1}} \ket{\epsilon_l}\otimes\ket{\epsilon_{l+1}}\otimes\ket{n}_{\color{blue}{s}}\;.
\end{equation}

The gauge-invariance constraint becomes therefore
\begin{equation}
\begin{split}
\label{eq:phys_constr}
\Psi^0_{\epsilon_{l},\epsilon_{l+1}}&(\epsilon_{l+1}-\epsilon_{l})=0\;,\\
\Psi^1_{\epsilon_{l},\epsilon_{l+1}}&(\epsilon_{l+1}-\epsilon_{l}-e_s)=0\;.
\end{split}
\end{equation}
A state belonging to the physically meaningful portion of the Hilbert space is then
\begin{equation}
\ket{\Psi^s_{l,l+1}} = \sum_{\epsilon_l}\sum_{n=0,1} \Psi^n_{\epsilon_{l}} \ket{\epsilon_l}\otimes\ket{\epsilon_{l}+ne_s}\otimes\ket{n}_{\color{blue}{s}}\;,
\end{equation}
where we recall that $e_s=1$ for a fermionic site and $e_s=-1$ for an anti-fermionic site. In order to obtain a useful encoded state for error correction, we can then apply a lowering operator $\hat{U}^\dag_{l+1}$ (for $e_s=1$) or a raising operator $\hat{U}_{l+1}$ (for $e_s=-1$) to the link $l+1$ controlled on the state of the site qubit. Calling this operation $\hat{W}_s$, for a site between an even and an odd link we have that
\begin{align}
\hat{W}_s\ket{\Psi^s_{l,l+1}}_{GL} &= \sum_{\epsilon_l}\sum_{n=0,1} \Psi^n_{\epsilon_{l}} \ket{\epsilon_l}_{GLE}\otimes\ket{\epsilon_{l}}_{GLO}\otimes\ket{n}_{\color{blue}{s}} \nonumber \\
= \sum_{\epsilon_l}&\sum_{n=0,1} \Psi^n_{\epsilon_{l}} \ket{\epsilon_l}\otimes\ket{\epsilon_l}\otimes\ket{\epsilon_{l}}\otimes\ket{n}_{\color{blue}{s}} \label{eq:g_to_l_map}
\end{align}
and the error correction procedure can be carried out on this new state. After one site has been processed, we apply the inverse $\hat{W}^\dag_s$ and move to the next site. In this approach, we effectively move from a gauge-invariant encoding that satisfies Eq.~\eqref{eq:phys_constr} to a logical encoding equivalent to the bit-flip code by using $\hat{W}_s$ and its inverse. This operation needs to be performed fault-tolerantly in order to ensure a controlled propagation of errors. In the next section we will restrict the discussion to the simpler case where the link registers are formed by a single qubit (ie. the flux cutoff is set to one). In that case, we show that the construction can be done in a fault-tolerant way. 

\begin{figure}
\[
    \Qcircuit @C=1em @R=1em {
        \lstick{\ket{\psi}_L} & \ctrl{1} & \qw & & & & \lstick{\ket{\psi}_L} & \ctrl{2} & \qw & \qw \\
        \lstick{\ket{\phi}_L} & \targ & \rstick{\ket{\psi \oplus \phi}_L} \qw & & & \push{\rule{2em}{0em}\equiv\rule{2em}{0em}} & \lstick{\ket{\phi}_L} & \qw & \ctrl{1} & \qw \\
        & & & & & & \lstick{\ket{0}} & \targ & \targ & \rstick{\ket{\psi \oplus \phi}} \qw 
}
\]
\label{fig:XOREquiv}
\caption{Equivalence for logical CNOT between logical qubits and for two logical-to-physical CNOTs.}
\end{figure}
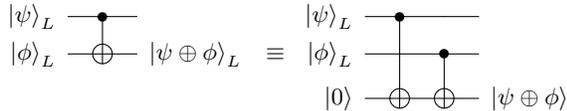

It is preferable to avoid directly implementing the $\hat{W}_s$ operation in practice as bit-flip errors can propagate from the site used as the control for the $l+1$ link. This can be done by adopting a full bit-flip encoding for the site alone and using the measurement of its stabilizers to correct for the induced error in the link. A more efficient strategy however is to compute the parity between the between the $l+1$ link and the adjacent site and store the result in a physical ancilla qubit, as shown schematically in \autoref{fig:XOREquiv}. The ancilla qubit can then be used as part of a logical encoding for the even-numbered links. We can catch errors of weight two or higher that occur from errors propagating past the logical CNOTs with the use of flag qubits (see \cite{Chao_2018}) and an example of this is shown in \autoref{fig:FTlogtophysCNOT}. This construction can be readily extended to situations with large cutoffs if we employ a unary encoding for the flux state using $m$ qubits per link, in which case we will require $m$ ancilla qubits. Generalizing the approach to larger cutoff values with various encodings is an important issue which will need to be addressed in future work (see also Sec.~\ref{sec:discussion}). As we show in more detail in the next section, using ancilla qubits to temporarily store the parity of the two links allows for a fault-tolerant error correction of a gauge theory with dynamical fermions using the unencoded sites directly, providing a significant saving in qubits.   

In summary, the number of link registers required for the Gauss' Law aided bit-flip error correction is $3N$ for a system with a total of $2N$ links. The total number of qubits required in a pure gauge simulation is therefore $9N + $ O(1) when restricting the unit electric cutoff case as done above and accounting for the additional $3$ physical qubits required for the phase-flip encoding. With the addition of dynamical fermions on $2N$ sites, one can exploit the relationship given by Gauss' law between the state of a site and its adjacent links to do error correction with a total of $15N + $ O(1) qubits, $9N$ for links and $6N$ for sites. These are impossible results if one instead attempts to perform error correction on individual qubits. The total space required would instead be $10N$ for the non-dynamical case and $20N$ for the dynamical one using the perfect $5$-qubit encoding from ~\cite{DiVincenzo1996,Laflamme1996}. As mentioned previously, better ratios are possible if one encodes multiple logical qubits at a time and we leave the extension to the more efficient encodings for future work~\cite{Gottesman1996,Calderbank1997}.

\section{Applications}

\label{sec:applications}

We specialize to a $\mathbb{Z}_2$ or truncated U(1) gauge theory in 1+1 dimensions with a cutoff of $\Lambda = 1$ for the flux in each link in this section. We work with a system with $2N$ sites and $2N$ links with PBCs. Site $k$, for an integer $k \geq 1$, is denoted by $S_k^1$ and the links coming into and out of the site are written as $L_k^1$ and $L_{k+1}^1$ respectively. 

Unless otherwise specified, every circuit in this section is a concatenated one, with every qubit encoded within the phase flip error-correcting code as in \autoref{fig:phasefliplogstate}. Logical qubits arising in this manner will be denoted with the subscript $L$. 

This section presents applications of the previous analysis to a few gauge theories in order of increasing generality. Beginning with the pure gauge theory case with no matter present on the sites, Gauss' Law gives the equality between the incoming and outgoing flux on a site. This allows us to use $\ket{L_k^1}$ and $\ket{L_{k+1}^1}$ as part of a single logical qubit for a concatenated bit-flip error correction procedure, where an additional qubit is introduced only if the links are even-numbered. 

This situation is easily generalized to a theory with static charges or non-dynamical fermions on the sites by applying a link lowering operator on an odd-numbered link conditioned on the classical state of the adjacent site. We then perform the error correction procedure on the appropriate logical link qubit before restoring the odd link's flux value with a raising operator. 

The scenario with dynamical fermions is dealt with by computing the parity between an odd-numbered link and its adjacent site and storing the result in a physical ancilla qubit. This qubit is then used as part of a logical encoding for the even-numbered links. Bit-flip checks can be performed on two sets of adjacent links and sites and the information from them can be used to correct errors on all the relevant qubits, provided there is a single bit-flip error per two such checks.

Note that we cannot perform this analysis entirely within the stabilizer formalism since our controlled operations are not logical ones at the level of the bit-flip encoding. While certain aspects of our treatment therefore rely on standard constructions within the stabilizer formalism, others involving Gauss' Law require us to move between physical and logical states. 

~
\\
~
\\
\textbf{Pure Gauge Theory}

~
\\
We adopt the notation $\ket{L_k^n}_L$ and $\ket{S_k^n}_L$ to denote link and site logical qubits respectively in the phase flip code, and $O^n_k$ for operators acting on the $k$-th logical link or site qubit. The case where $n=1$ in the superscript denotes the original logical qubit and logical qubit operator, while values of $n$ greater than $1$ denote the auxiliary link or site qubits introduced for the error correction procedures. If a given logical site or link is the only one of its kind, we will drop the superscript. 

In each of the subsequent circuits, we introduce no additional qubits for the odd numbered links and an auxiliary qubit $\ket{L^{2}_{2k}}_L$ for each even numbered link $\ket{L^1_{2k}}_L$ that is arranged to have the same state as $\ket{L_{2k}^1}_L$. The constraint that the fluxes satisfy Gauss' Law at site $\ket{S_{2k}^1}_L$ with associated link qubits $\{\ket{L_{2k}^1}_L, \ket{L_{2k}^2}_L,\ket{L_{2k+1}}_L \}$ will then ensure that $\ket{L_{2k}^1}_L = \ket{L_{2k}^2}_L = \ket{L^{1}_{2k+1}}_L$, or more precisely that the wave-function factorizes as in Eq.~\eqref{eq:gauss_logical_state_pg} above.

\begin{figure}[t]
    \[
            \qquad\qquad\Qcircuit @C=1em @R=1em {
            \lstick{\ket{L_{2k}^1}_L} & \ctrl{3} & \qw & \qw & \qw & \qw & \gate{X} & \qw & \qw & \qw \\
            \lstick{\ket{L^{2}_{2k}}_L} & \qw & \ctrl{2} & \ctrl{3} & \qw & \qw & \qw & \gate{X} & \qw & \qw \\
            \lstick{\ket{L_{2k+1}}_L} & \qw & \qw & \qw & \ctrl{2} & \qw & \qw & \qw & \gate{X} & \qw \\   
            \lstick{\ket{0}} & \targ & \targ & \qw & \qw & \meter & \cctrl{-3} & \cctrl{-2} & \cctrlo{-1} \\
            \lstick{\ket{0}} & \qw & \qw & \targ & \targ & \meter & \cctrlo{-1} & \cctrl{-1} & \cctrl{-1} 
            }
    \]
\caption{Bit-flip syndrome measurement and correction operations for links. The subscript $L$ denotes the phase-flip encoding of these qubits shown in \autoref{fig:phasefliplogstate}. Each CNOT operation here consists of three individual CNOTs from the qubits in the underlying phase-flip encoding. Here $X$ is the operation $X \otimes X \otimes X$ which acts as a logical $X$.}
\label{fig:puregaugeEC} 
\end{figure}
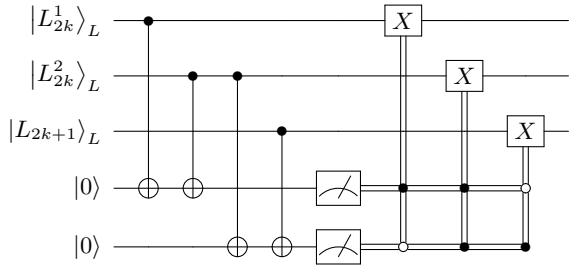

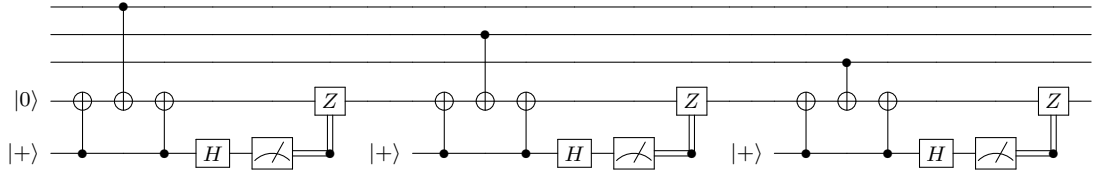
\begin{figure*}
\footnotesize
\[
    \Qcircuit @C=1em @R=1em {
    \lstick{} & \qw & \ctrl{3} & \qw & \qw & \qw & \qw & \qw & \qw & \qw & \qw & \qw & \qw & \qw & \qw & \qw & \qw & \qw & \qw & \qw& \qw & \qw & \qw & \qw & \qw & \qw \\
    \lstick{} & \qw & \qw & \qw & \qw & \qw & \qw & \qw & \qw & \qw& \qw & \ctrl{2} & \qw & \qw & \qw & \qw & \qw & \qw & \qw  & \qw & \qw& \qw & \qw & \qw & \qw & \qw \\
    \lstick{} & \qw & \qw & \qw & \qw & \qw & \qw & \qw & \qw & \qw & \qw & \qw & \qw & \qw & \qw & \qw & \qw & \qw & \qw & \qw & \ctrl{1}  & \qw & \qw & \qw & \qw & \qw \\
    \lstick{\ket{0}} & \targ & \targ & \targ & \qw & \qw & \gate{Z} & \qw & \qw & \qw & \targ & \targ & \targ & \qw & \qw & \gate{Z} & \qw & \qw & \qw & \targ & \targ & \targ & \qw & \qw & \gate{Z} & \qw \\
    \lstick{\ket{+}} & \ctrl{-1} & \qw & \ctrl{-1} & \gate{H} & \meter & \cctrl{-1} & & & \lstick{\ket{+}} & \ctrl{-1} & \qw & \ctrl{-1} & \gate{H} & \meter & \cctrl{-1} & & & \lstick{\ket{+}} & \ctrl{-1} & \qw & \ctrl{-1} & \gate{H} & \meter & \cctrl{-1}   
    }
\]
\caption{Fault tolerant logical-to-physical CNOT using flag qubits.}
\label{fig:FTlogtophysCNOT}
\end{figure*}

As noted in the previous section, the grouping $\{\ket{L_{2k}^1}_L,\ket{L_{2k}^2}_L, \ket{L_{2k+1}}_L\}$ acts as a 3-qubit logical encoding for the link logical qubit $\ket{L_{2k}^1}_L$ that can be used in the bit flip error correction code outlined in \autoref{fig:puregaugeEC}. The circuit is however sensitive to single qubit phase flip errors that propagate to weight two or three phase flip errors. This is due to the use of logical to physical CNOT gates, which enable the propagation of physical $Z$ errors from the target of these CNOTs to logical $Z$ errors on the links. We therefore employ the fault-tolerant logical-to-physical CNOT depicted in \autoref{fig:FTlogtophysCNOT}. This CNOT is implicitly used in all the subsequent circuits where logical-to-physical CNOT operations occur.

There is an alternative way to accomplish this error correction procedure without introducing an additional logical qubit for the even numbered links. This method simply requires doing overlapping Gauss's law checks on the qubits $\{\ket{L_{2k}}_L, \ket{L_{2k+1}}_L, \ket{L_{2k+2}}_L\}$ and the qubits in \autoref{fig:puregaugeEC} can then be replaced with these respectively. The number of qubits required to fully error correct a 1+1 lattice system with $2N$ links would then be $6N$ instead of $9N$. However, we can then only tolerate at most one error occurring over a larger portion of the lattice, namely a group of three consecutive links, compared to the previous scenario where we could tolerate at most one error over the more localized grouping $\{\ket{L_{2k}^1}_L,\ket{L_{2k}^2}_L, \ket{L_{2k+1}}_L\}$. Thus there is a trade-off between the number of qubits needed for our compressed encoding and the spatial region over which the Gauss's law checks need to be performed to localize a bit-flip error. This pattern will recur in the subsequent cases we consider and becomes less favorable in larger dimensions.  

~
\\
\textbf{Non-Dynamical Fermions}

~
\\
It is straightforward to generalize the results of the preceding subsection to the case where we have non-dynamical fermions on the sites. Since we know in this case whether or not a certain physical site $S_{2k}^1$ (this is not a logical site qubit but a classical binary variable) has a fermion, it is not necessary to introduce an ancilla qubit and use the equivalence in \autoref{fig:XOREquiv}. The controlled operation $\hat{W}_s$ introduced above (see Eq.~\eqref{eq:g_to_l_map}) is instead replaced by its classically controlled counterpart. As such, we apply a classically conditioned link-lowering operator to $\ket{L_{2k+1}}_L$ to ``lower" its flux value and have it match that of $\ket{L_{2k}^1}_L$. We then apply the preceding bit-flip error correction code and restore the flux value of $\ket{L_{2k+1}^1}_L$ back to its original value with a classically conditioned link-raising operation.

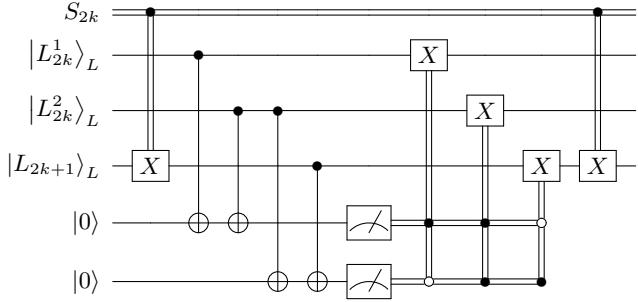
\begin{figure}[t]
    \[\qquad\qquad
        \Qcircuit @C=0.8em @R=1em {
            \lstick{S_{2k}} & \cctrl{3} & \cw & \cw & \cw & \cw & \cw & \cw & \cw & \cw & \cctrl{3} & \cw \\
            \lstick{\ket{L_{2k}^1}_L} & \qw & \ctrl{3} & \qw & \qw & \qw & \qw & \gate{X} & \qw & \qw & \qw & \qw \\
            \lstick{\ket{L^{2}_{2k}}_L} & \qw & \qw & \ctrl{2} & \ctrl{3} & \qw & \qw & \qw & \gate{X} & \qw & \qw & \qw \\
            \lstick{\ket{L_{2k+1}}_L} & \gate{X} & \qw & \qw & \qw & \ctrl{2} & \qw & \qw & \qw & \gate{X} & \gate{X} & \qw \\   
            \lstick{\ket{0}} & \qw & \targ & \targ & \qw & \qw & \meter & \cctrl{-3} & \cctrl{-2} & \cctrlo{-1} & & \\
            \lstick{\ket{0}} & \qw & \qw & \qw & \targ & \targ & \meter & \cctrlo{-1} & \cctrl{-1} & \cctrl{-1} & &
        }
    \]
\caption{Syndrome measurement and correction operations for LGT system with non-dynamical fermions. The circuit holds for arbitrary sites $S_k$ and not just even numbered ones. Note that the circuit can be simplified by using commutation relations to move the initial classically controlled $X$ to the end of the circuit, thereby replacing the classical controls with a classically controlled flip of the last measurement outcome.}\label{fig:ECnondynamical} 
\end{figure}

Given our assumptions on the flux cutoff, we have the circuit in \Cref{fig:ECnondynamical} for the non-dynamical fermionic case. This construction also holds for an arbitrary site $S_k$ and not just even numbered ones. Note that the circuit can be expressed in a simpler form by exploiting commutation relations to move the initial classically controlled $X$ to the end of the circuit. This effectively replaces this operation with a classically controlled flip of the last measurement outcome. 

As in the pure gauge theory case, we can dispense with the extra logical qubit $\ket{L^2_{2k}}_L$ and extend the Gauss's law check to the group of qubits $\{\ket{L_{2k}^1}_L,\ket{L_{2k}^2}_L, \ket{L_{2k+1}}_L\}$, replacing the link qubits in \autoref{fig:ECnondynamical} accordingly. The number of qubits needed for this approach is again $6N$ like the pure gauge case instead of $9N$ with the above encoding, and the trade-offs are identical as well. 

~
\\ 
\textbf{Dynamical Fermions}

~
\\
A na\"{i}ve approach to extend the preceding constructions to the case with dynamical staggered fermions on the sites is to apply a CNOT gate to $\ket{L_{2k+1}}_L$ conditioned on the state of $\ket{S_{2k}}_L$. We could then introduce additional qubits $\ket{S^2_{2k}}_L$ and $\ket{S^3_{2k}}_L$ to create a bit-flip encoding of $\ket{S_{2k}}_L$ and thereby protect it from bit-flip errors. Errors propagating to $\ket{L_{2k+1}}_L$, $\ket{L^1_{2k}}_L$, and $\ket{L^2_{2k}}_L$ can be corrected with a separate bit-flip code involving these qubits, with additional recovery operations included to account for the propagation of $X$ errors past the control of the CNOT.

This construction only involves Clifford operations and can easily be made fault tolerant. It however requires $18N+6N+3N=27N$ qubits for an $2N$ site and $2N$ link system excluding ancillas, which is suboptimal compared to the $20N$ qubits obtained from using the $[5,1,3]$ encoding for each site and link.

We instead develop a procedure that further exploits Gauss' law to reduce the cost to $15N$ qubits. This is achieved by noticing that, as discussed in Sec.~\ref{subsec:gauss_construction}, given the state on a site $S_{2k+1}$ and its adjacent links $L_{2k+1}$ and $L_{2k+2}$, a bit-flip error on any of these qubits will cause a violation of Gauss' law. The ambiguity in where the error occurred can be resolved by overlapping Gauss' law checks on the preceding and subsequent site and their adjacent links. In fact, we will show only two such checks are needed to resolve the ambiguity. 

\begin{figure*}
\[
    \quad\quad\quad\quad\;\;\Qcircuit @C=0.75em @R=1em {
    \lstick{} & & & & & & \mathcal{G}_{2k+1} \gategroup{2}{2}{11}{12}{1.7em}{--} & & & & & & & & & & & & & & & \mathcal{G}_{2k} \gategroup{2}{17}{11}{27}{1.7em}{--} & & & & & \\
    \lstick{\ket{S_{2k}}_L}     & \qw    & \qw    & \qw    & \qw    & \qw    & \qw    & \qw    & \qw     & \qw     & \qw       & \qw       & \qw                  & \qw & \qw & \qw                                   &\ctrl{7}& \qw    & \qw    & \qw    & \qw    & \qw    & \qw & \ctrl{7} & \qw & \qw & \qw & \gate{X} & \qw & \qw & \qw \\
    \lstick{\ket{S_{2k+1}}_L}   &\ctrl{6}& \qw    & \qw    & \qw    & \qw    & \qw    & \qw    &\ctrl{6} & \qw     & \qw       & \qw       & \qw                  & \qw & \qw & \qw                                   & \qw    & \qw    & \qw    & \qw    & \qw    & \qw    & \qw & \qw & \qw & \qw & \qw & \qw & \gate{X} & \qw & \qw \\
    \lstick{\ket{L_{2k}^1}_L}   & \qw    & \qw    & \qw    & \qw    & \qw    & \qw    & \qw    & \qw     & \qw     & \qw       & \qw       & \qw                  & \qw & \qw & \qw                                   & \qw    & \qw    &\ctrl{6}& \qw    & \qw    & \qw    & \qw & \qw & \qw & \gate{X} & \qw & \qw & \qw & \qw & \qw \\
    \lstick{\ket{L_{2k}^2}_L}   & \qw    & \qw    & \qw    & \qw    & \qw    & \qw    & \qw    & \qw     & \qw     & \qw       & \qw       & \qw                  & \qw & \qw & \qw                                   & \qw    & \qw    & \qw    &\qw     &\qw     &\ctrl{6}& \qw & \qw & \qw & \qw & \gate{X} \push{\rule{1em}{0em}} & \qw & \qw & \qw & \qw \\
    \lstick{\ket{L_{2k+1}}_L}   & \qw    &\ctrl{3}& \qw    & \qw    & \qw    & \qw    &\ctrl{3}& \qw     & \qw     & \qw       & \qw       & \qw                  & \qw & \qw & \qw                                   & \qw    &\ctrl{3}& \qw    & \qw    & \qw    & \qw    & \ctrl{3} & \qw & \qw & \qw & \qw & \qw & \qw & \gate{X} & \qw \\
    \lstick{\ket{L_{2k+2}^1}_L} & \qw    & \qw    &\ctrl{3}& \qw    & \qw    & \qw    & \qw    & \qw     & \qw     & \gate{X}  & \qw       & \qw                  & \qw & \qw & \qw                                   & \qw    & \qw    & \qw    & \qw    & \qw    & \qw    & \qw & \qw & \qw & \qw & \qw & \qw & \qw & \qw & \qw \\
    \lstick{\ket{L_{2k+2}^2}_L} & \qw    & \qw    & \qw    & \qw    & \qw    &\ctrl{3}& \qw    & \qw     & \qw     & \qw       & \gate{X}  & \qw                  & \qw & \qw & \qw                                   & \qw    & \qw    & \qw    & \qw    & \qw    & \qw    & \qw & \qw & \qw & \qw & \qw & \qw & \qw & \qw & \qw \\
    \lstick{\ket{0}}          & \targ  & \targ  & \qw    &\ctrl{1}&\ctrl{2}& \qw    & \targ  & \targ   & \qw     & \qw       & \qw       &                      &     &     &\lstick{\ket{0}}                     & \targ  & \targ  & \qw    &\ctrl{1}&\ctrl{2}&\qw     & \targ & \targ & \qw & \qw & \qw & \qw & \qw & \qw & \qw \\
    \lstick{\ket{0}}            & \qw    & \qw    & \targ  & \targ  & \qw    & \qw    & \meter &\cctrl{2}& \cw     & \cctrl{-3}&\cctrlo{-2}&                      &     &     & \lstick{\ket{0}}                      & \qw    & \qw    & \targ  & \targ  & \qw    & \qw    & \meter & \cw & \cw & \cctrl{-6} & \cctrlo{-5} & \cctrl{-8} & \cctrlo{-7} & \cctrl{-4} & \\
    \lstick{\ket{0}}            & \qw    & \qw    & \qw    & \qw    & \targ  & \targ  & \meter & \cw     &\cctrl{2}&\cctrlo{-1}& \cctrl{-1}&\push{\rule{0em}{1em}}&     &     &\lstick{\ket{0}} \push{\rule{0em}{1em}}& \qw    & \qw    & \qw    & \qw    & \targ  & \targ  & \meter & \cw & \cw & \cctrlo{-1} & \cctrl{-1} & \cctrl{-1} & \cctrlo{-1} & \cctrl{-1}  & \\
     &  &  &  &  &  &  &  &  & \cw & \cw & \cw & \cw & \cw & \cw & \cw & \cw & \cw & \cw & \cw & \cw     & \cw & \cw & \cw & \cw & \cw & \cw & \cctrlo{-1} & \cctrl{-1} & \cctrl{-1} &  \\
     &  &  &  &  &  &  &  &  &  & \cw & \cw & \cw & \cw & \cw & \cw \cw& \cw & \cw & \cw & \cw & \cw &\cw &\cw &\cw &\cw & \cw &\cw &\cctrlo{-1} & \cctrl{-1} & \cctrl{-1} &  
    }
\]
\caption{Circuit for using information from Gauss' Law checks $\mathcal{G}_{2k}$ and $\mathcal{G}_{2k+1}$ on $S_{2k}$, $S_{2k+1}$, and their adjacent links to correct bit-flip errors on them. Bit-flip errors on even-numbered links can be corrected using the information obtained from one such check. Bit-flip errors on sites and odd-numbered links require information from two such checks to be corrected. See \autoref{tab:syndresultsdynamical} for the corresponding syndrome outcomes.}
\label{fig:GaussLawDynamical}
\end{figure*}
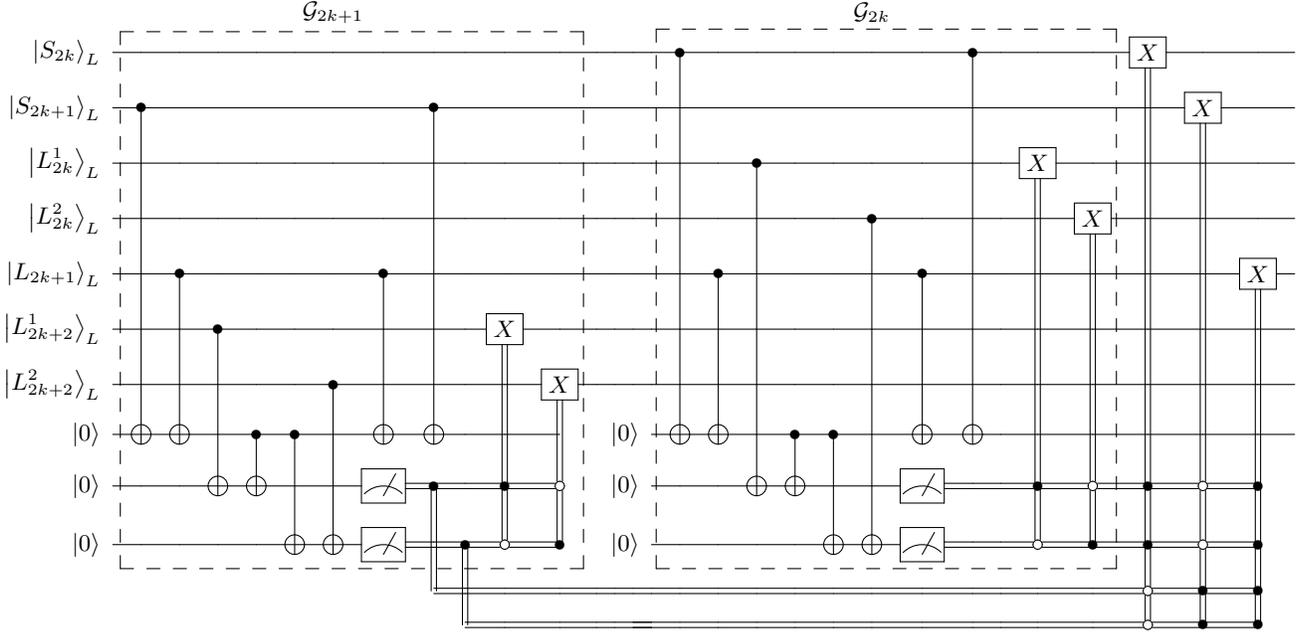

\autoref{fig:GaussLawDynamical} depicts to overlapping Gauss' law checks $\mathcal{G}_{2k}$ and $\mathcal{G}_{2k+1}$ on sites $S_{2k}$ and $S_{2k+1}$ respectively. Consider $\mathcal{G}_{2k+1}$ in particular. We can compute the parity between $\ket{S_{2k+1}}_L$ and $\ket{L_{2k+1}}_L$ as per \autoref{fig:XOREquiv} and store this result into a physical ancilla qubit $\ket{0}$, which serves as a proxy for $\ket{L_{2k+1}}_L$ but with its flux value reset. We can then treat $\ket{0}$, $\ket{L_{2k+2}^1}_L$, $\ket{L^2_{2k+2}}_L$ as a logical qubit for a bit-flip error correction code and uncompute the value in $\ket{0}$ afterwards. These steps constitute $\mathcal{G}_{2k+1}$ for the odd sites and their neighboring links and are repeated analogously across the lattice. 

To see how the information obtained from the stabilizer measurements employing the last two ancilla qubits in \autoref{fig:GaussLawDynamical} helps us correct errors on the data qubits, consider $\mathcal{G}_{2k+1}$ first. The syndrome measurement outcomes $10$ and $01$ uniquely identify bit-flip errors occurring on the even links $\ket{L_{2k+2}^1}_L$ and $\ket{L^2_{2k+2}}_L$ respectively. The remaining non-trivial outcome $11$ indicates a bit-flip error on either $\ket{L_{2k+1}}_L$ or $\ket{S_{2k+1}}_L$ with no way to resolve the ambiguity.

If the error occurred on $\ket{S_{2k+1}}_L$ however, we would obtain the syndrome $00$ from the check $\mathcal{G}_{2k}$ as Gauss' law is satisfied between $S_{2k}, L_{2k}$, and $L_{2k+1}$. If the error occurred on $\ket{L_{2k+1}}_L$, we would then obtain the syndrome $11$ from $\mathcal{G}_{2k}$. As these two cases yield distinct syndromes, we can resolve the aforementioned ambiguity as shown in \autoref{tab:syndresultsdynamical}. Note that this analysis will only hold if a bit-flip error occurs once during the checks $\mathcal{G}_{2k}$ and $\mathcal{G}_{2k+1}$. 
 
\begin{table}[b]
\centering 
\begin{tabular}{| c | c | c|} 
\hline
$\mathcal{G}_{2k+1}$ & $\mathcal{G}_{2k}$ & Error Location  \\ 
\hline
00 & 00 & None \\
\hline
00 & 11 & $\ket{S_{2k}}_L$ \\
\hline 
11 & 00 & $\ket{S_{2k+1}}_L$ \\ 
\hline
11 & 11 & $\ket{L_{2k+1}}_L$ \\
\hline 
\end{tabular}
\caption{Table showing how to resolve the ambiguity in the location of a bit-flip error associated to the syndrome $11$ for $\mathcal{G}_{2k+1}$ using the syndrome results of $\mathcal{G}_{2k}$. See \autoref{fig:GaussLawDynamical}.}
\label{tab:syndresultsdynamical}
\end{table}

To assess the fault-tolerance of this construction, we first consider the possibility that single-qubit bit flip errors occurring at the physical level can propagate to undetectable errors at level of the phase-flip code. Consider WLOG an error of the form $X \otimes I \otimes I$ occurring at the level of the underlying phase-flip code. On the logical codespace defined by the basis $\ket{0}_L = (\ket{+++} + \ket{---})/\sqrt{2}$ and $\ket{1}_L = (\ket{+++} - \ket{---})/\sqrt{2}$, this error maps $\ket{0}_L \mapsto \ket{1}_L$. In other words, a physical bit-flip error propagates to a logical bit-flip error which can be detected by our bit-flip correction code. It is easily seen that two bit flip errors leave our code space invariant. The circuit's sensitivity to single qubit phase flip errors propagating to weight two or three phase flip errors from the logical-to-physical CNOT gates is again resolved with the fault-tolerant CNOT construction shown in \autoref{fig:FTlogtophysCNOT}.

Logical bit-flip errors occurring on the logical qubits only propagate to the ancilla qubits and not to the data qubits. The possibility of more than one such error occurring on the data qubits is suppressed by higher powers of $p$, where $p$ is the probability of an error occurring. 

\begin{figure}
\includegraphics[width=0.49\textwidth]{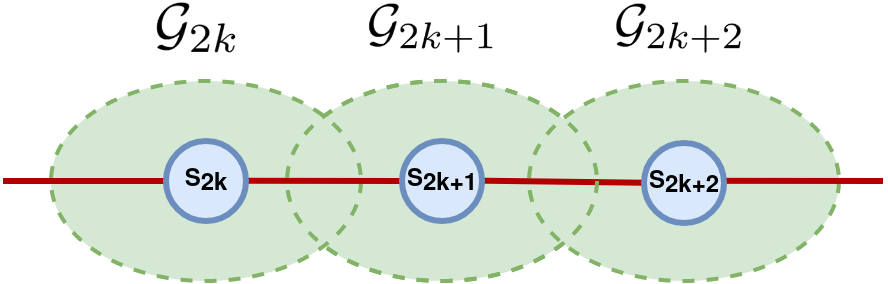}
\caption{Schematic illustration of how one can correct any single-qubit error on $S_{2k+1}$ and its adjacent logical link qubits by 3 overlapping Gauss's law checks. This requires that only one single-qubit error occurs over the links and qubits involved in the 3 Gauss's law checks.  \label{fig:GL3sites}}
\end{figure}

We can again consider the situation where we dispense with the extra qubits for the even-numbered links and correct bit-flip errors solely based on information obtained from Gauss's law. This modified situation is depicted in \autoref{fig:GL3sites} and shows how three overlapping Gauss's law checks can be used to detect an error on $\ket{S_{2k+1}}$ or its adjacent links. Note that this requires only one single-qubit error to occur on the links and qubits involved in the checks, and so we can lessen the memory overhead required for error-correction in this setting at the expense of a worse error tolerance over a larger region of the lattice. This modified approach then requires $3(2N) + 3(2N) = 12N$ qubits for a 1+1 dimensional lattice system with $2N$ staggered fermions and $2N$ links. The corresponding circuit is similar to \autoref{fig:GaussLawDynamical}, with the exception that the extra even-numbered link qubits and intermediate classical recovery operations in each $\mathcal{G}_{j}$ are removed and the logical qubits $\ket{S_{2k+2}}_L, \ket{S_{2k+3}}_L$ are present. We simply perform the parity checks $\mathcal{G}_{2k}, \mathcal{G}_{2k+1}, \mathcal{G}_{2k+2}$, store the results into 3 ancilla qubits, and use the outcomes to correct the error that happened on $\ket{S_{2k+1}}_L, \ket{L_{2k+1}}_L, \ket{L_{2k+2}}_L$. \autoref{tab:syndresultsdynamextend} depicts the possible outcomes from these checks and shows that we can localize the error. 

\begin{table}[b]
\centering 
\begin{tabular}{| c | c | c | c |} 
\hline
$\mathcal{G}_{2k}$ & $\mathcal{G}_{2k+1}$ & $\mathcal{G}_{2k+2}$ & Error Location  \\ 
\hline
0 & 0 & 0 & None \\
\hline
0 & 1 & 1 & $\ket{L_{2k+2}}_L$ \\
\hline
0 & 1 & 0 & $\ket{S_{2k+1}}_L$ \\
\hline 
1 & 1 & 0 & $\ket{L_{2k+1}}_L$ \\ 
\hline
\end{tabular}
\caption{Table showing how to localize the error occurring on $\ket{S_{2k+1}}_L$ or its adjacent logical link qubits $\ket{L_{2k+1}}_L, \ket{L_{2k+2}}_L$ by overlapping Gauss's law checks. Each error is associated with a unique syndrome, enabling error correction. See \autoref{fig:GL3sites} as well.}
\label{tab:syndresultsdynamextend}
\end{table}

\section{Extension to two dimensions}
\label{sec:2dlgt}

\begin{figure}
\includegraphics[width=0.2\textwidth]{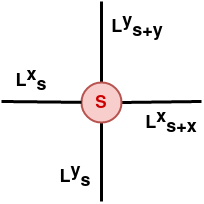}
\caption{Convention for the indexing of the link variables around a fermionic site ${\bf s}=(x_s,y_s)$. \label{fig:2dsite}}
\end{figure}

\begin{figure*}
\includegraphics[width=0.9\textwidth]{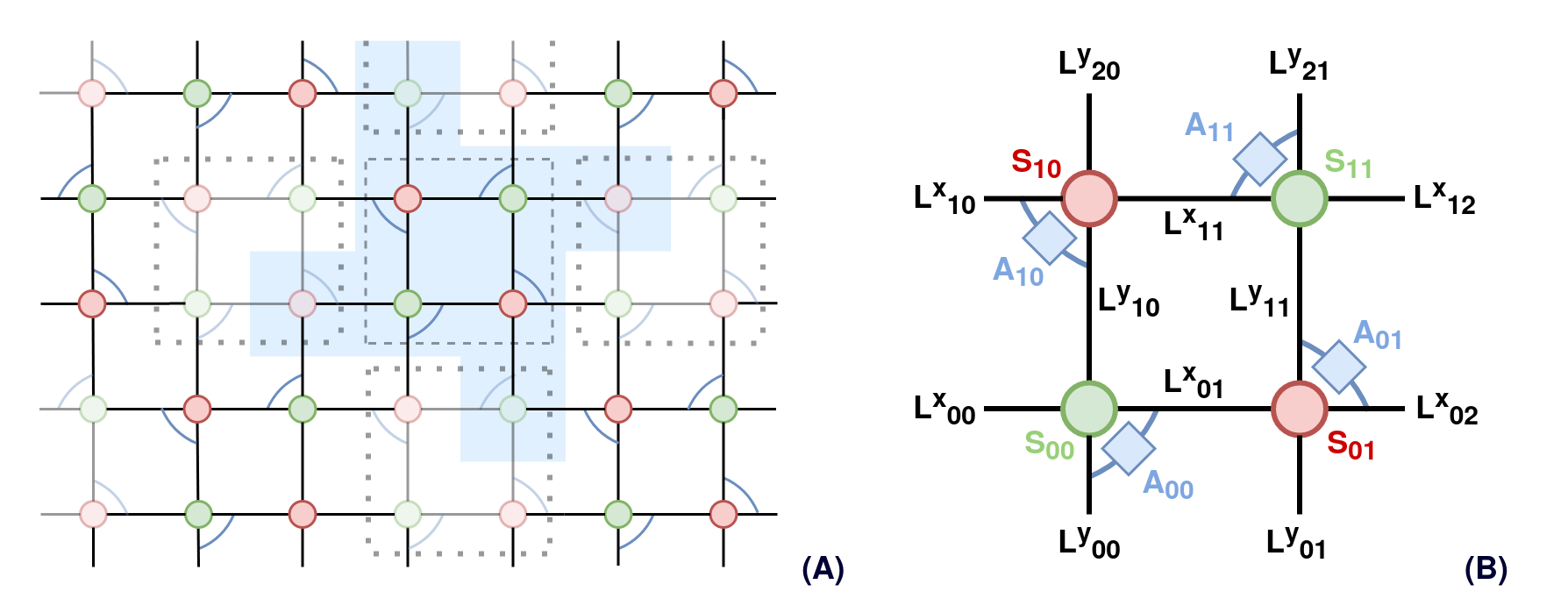}
\caption{Panel (A) shows the adopted bi-partition of the 2D lattice into plaquettes used for the two rounds of the error correction procedure: the first round acts on bold links and sites (e.g. the region delimited by the dashed line) the second round acts on the light colored links and sites and the 8 external links attached to them (e.g. the regions delimited by dotted lines). The blue shaded are shown in panel (A) is the patch of the 2D lattice where one bit flip error can occur and be corrected. Panel (B) shows in detail the structure of a region interested by one round of the error correction procedure.} 
\label{fig:2dlattice}
\end{figure*}

The scheme described in the previous sections can be also extended to higher spatial dimensions. In order to give a concrete example we describe here how one could implement a compressed error correcting code for a simple 2+1 dimensional lattice gauge theory. We consider in detail the case of $\bb{Z}_2$ but the construction can be also applied to U(1). We start from a different form of Gauss' law which allows for a more direct connection to the stabilizer formalism. The Gauss' operator at each site ${\bf s}=(x_s,y_s)$ and it's neighboring links is defined as
\begin{equation}
\label{eq:2dgauss}
\hat{G}({\bf s}) = Z^x_{\bf s}Z^y_{\bf s}Z^x_{{\bf s}+x}Z^y_{{\bf s}+y}(-1)^{ \hat{q}({\bf s})}\;,
\end{equation}
and physical states satisfy the condition
\begin{equation}
\hat{G}({\bf s}) \ket{\Psi}_{phys} = \ket{\Psi}_{phys}\;.
\end{equation}
In the expression Eq.~\eqref{eq:2dgauss} we use the indexing convention for the links as displayed in Fig.~\ref{fig:2dsite} and the charge operator using staggered fermions is given by
\begin{equation}
\hat{q}({\bf s}) = \psi^\dagger({\bf s})\psi({\bf s})-\frac{1}{2}\left(1-(-1)^{x_s+y_s}\right),
\end{equation}
with $\hat{\psi}^\dagger({\bf s})$ and $\hat{\psi}({\bf s})$ the fermionic creation and annihilation operators at site ${\bf s}$.
With this definition, at even site the states $\{\ket{0},\ket{1}\}$ indicate the absence/presence of a fermion of charge $1$ while for odd sites the role is reversed and $\{\ket{0},\ket{1}\}$ represent the presence/absence of an anti-fermion of opposite charge. In the following we will denote fermionic sites with a red color (as in Fig.~\ref{fig:2dsite}) and anti-fermion sites with a green color. The convention for the indexing of links is unaffected. The presence of static charges at different sites is easily included by adding their contributions in the charge at the site. In order to keep the current exposition simple we neglected this possibility.

Using the Pauli $Z$ operator at the site we can express the Gauss law operator more explictly as
\begin{equation}
\hat{G}({\bf s}) = (-1)^{x_s+y_s}Z^x_{\bf s}Z^y_{\bf s}Z^x_{{\bf s}+x}Z^y_{{\bf s}+y}Z_{\bf s}\;,
\end{equation}
and we see then that physical states are eigenvectors of 
\begin{equation}
\label{eq:2dstab}
Z^x_{\bf s}Z^y_{\bf s}Z^x_{{\bf s}+x}Z^y_{{\bf s}+y}Z_{\bf s}
\end{equation}
with positive eigenvalue on even sites and negative eigenvalue on odd sites. As alluded above, this form is reminiscent of the stabilizer formalism.

\begin{table*}[]
\begin{tabular}{c|c|c||c|c|c||c|c|c||c|c|c||l}
$P^a_{(0,0)}$ & $P^b_{(0,0)}$ & $P^c_{(0,0)}$ & $P^a_{(0,1)}$ & $P^b_{(0,1)}$ & $P^c_{(0,1)}$ & $P^a_{(1,1)}$ & $P^b_{(1,1)}$ & $P^c_{(1,1)}$ & $P^a_{(1,0)}$ & $P^b_{(1,0)}$ & $P^c_{(1,0)}$ & error location \\ \hline
+ & + & + & + & - & - & + & + & + & + & - & - & none \\ \hline
+ & + & + & + & - & - & + & + & + & - & - & + & $A_{10}$\\ \hline
+ & + & + & + & - & - & - & + & - & + & - & - & $A_{11}$ \\ \hline
+ & + & + & - & - & + & + & + & + & + & - & - & $A_{01}$ \\ \hline
- & + & - & + & - & - & + & + & + & + & - & - & $A_{00}$ \\ \hline
+ & + & + & + & - & - & + & + & + & - & + & - & $L^x_{10}$\\ \hline
+ & + & + & + & - & - & - & - & + & + & - & - & $L^y_{21}$\\ \hline
+ & + & + & - & + & - & + & + & + & + & - & - & $L^x_{02}$\\ \hline
- & - & + & + & - & - & + & + & + & + & - & - & $L^y_{00}$\\ \hline
+ & - & - & + & - & - & + & + & + & - & + & - & $L^y_{10}$\\ \hline
+ & + & + & + & - & - & - & - & + & + & + & + & $L^x_{11}$\\ \hline
+ & + & + & - & + & - & - & - & + & + & - & - & $L^y_{11}$\\ \hline
- & - & + & + & + & + & + & + & + & + & - & - & $L^x_{01}$\\ \hline
+ & + & + & + & - & - & + & + & + & + & + & + & $s_{10}$, $L^y_{20}$\\ \hline
+ & + & + & + & - & - & + & - & - & + & - & - & $s_{11}$, $L^x_{12}$\\ \hline
+ & + & + & + & + & + & + & + & + & + & - & - & $s_{01}$, $L^y_{01}$\\ \hline
+ & - & - & + & - & - & + & + & + & + & - & - & $s_{00}$, $L^x_{00}$ \\ \hline
\end{tabular}
\caption{Syndrome measurements outcomes, and corresponding error location, for the 12 stabilizers used on one complete plaquette. The labeling conventions follow Fig.~\ref{fig:2dlattice} (B). We have excluded outcomes corresponding to more than one error.}\label{tab:2dparitycheck}
\end{table*}

The error correcting procedure proceeds as follow: (1) we first partition the full lattice into square plaquettes and their neighboring links and proceed in two rounds by first considering the plaquettes indicated with bold colors in panel (A) of Fig.~\ref{fig:2dlattice} (see e.g. the one indicated by the dashed line) and in the second round we consider the second half of the plaquettes denoted with light colors (for the highlighted plaquette, these correspond to those indicated by dotted lines); (2) for each individual plaquette we perform a sequence of parity checks. Before describing these, note that we require an additional qubit for every pair of links denoted as a blue curved line in the diagrams of Fig.~\ref{fig:2dlattice}. These qubits are used to store the parity of the two links they touch, for instance
\begin{equation}
Z^x_{(1,0)}Z^y_{(1,0)} \ket{\Phi} = Z_{A_{10}}\ket{\Phi}\;,
\end{equation}
where with $\ket{\Phi}$ we denote the full state. This can be achieved by simply applying two CNOT gates with control on the links and target on the ancillary qubit initially set in $\ket{0}$. Then, whenever we modify the state of a link, we correspondingly update the state of the ancilla by applying another CNOT. These are CNOT operations between phase-flip encoded qubits and can be done transversally. At this point three separate parity checks are performed on each site participating in the plaquette and its four connected links. Starting from the site $(0,0)$ on the bottom left, we have the stabilizers
\begin{equation}
\begin{split}
P^a_{(0,0)} &= Z^y_{(0,0)}Z^x_{(0,1)}Z_{A_{00}}\\
P^b_{(0,0)} &= Z^y_{(0,0)}Z^x_{(0,1)}Z^y_{(1,0)}Z^x_{(0,0)}Z_{(0,0)}\\
P^c_{(0,0)} &= Z_{A_{00}}Z^y_{(1,0)}Z^x_{(0,0)}Z_{(0,0)}\;,
\end{split}
\end{equation}
where $Z_{(0,0)}$ is the operator acting on the qubit at the site. The parity check table for one site is reported in Tab.~\ref{tab:2d1site}, together with the possible error location. We can see that only 4 outcomes correspond to either one error or no errors. In the following discussion on the complete plaquette we restrict our attention to these situations only. Note however that errors on the ancillas $A_{\bf s}$ can be uniquely determined by only stabilizer measurements on sites and we could therefore detect and fix an arbitrary number of them.

We will use the same convention for all the other sites and take $P^a_{\bf s}$ to be the weight 3 stabilizer between the ancilla $A_{\bf s}$ and the two link connected to it, $P^b_{\bf s}$ the weight 5 stabilizer from the Gauss' law operator Eq.~\eqref{eq:2dstab} and $P^c_{\bf s}$ the weight 4 stabilizer obtained applying Gauss' law to the ancilla $A_{\bf s}$, the two sites not connected to it and the site. Accounting for the difference between the correct eigenspace of the Gauss'law stabilizer for even (fermionic) and odd (anti-fermionic) sites, the parity check table for a complete plaquette is reported in Tab.~\ref{tab:2dparitycheck}. Note that the last three syndrome outcomes do not allow to uniquely localize the error as it could have happened on either the site or the external link opposite to the location of the ancilla $A_{\bf s}$. Similarly to the construction described above for the one dimensional case, this ambiguity can be completely resolve by measuring one stabilizer on the neighboring plaquettes. In particular, under the assumption that only one error can occur in the shaded area in panel (A) of Fig.~\ref{fig:2dlattice} during the two rounds of check, it is enough to measure the $P^a$ stabilizers on the four highligthed sites. For completeness, we report the full syndrome measurement table required to distinguish these cases in Tab~\ref{tab:2dparitycheck_nb}.
\begin{table}[b]
\begin{tabular}{c|c|c|l}
$P^a_{(0,0)}$ & $P^b_{(0,0)}$ & $P^c_{(0,0)}$ & error location\\ \hline
+ & + & + & none \\ \hline
+ & + & - & 2+ \\ \hline
+ & - & + & 2+ \\ \hline
+ & - & - & $s_{00}$,$L^x_{00}$,$L^y_{10}$ \\ \hline
- & + & + & 2+ \\ \hline
- & + & - & $A_{00}$ \\ \hline
- & - & + & $L^y_{00}$, $L^x_{01}$ \\ \hline
- & - & - & 2+
\end{tabular}
\caption{Syndrome measurement outcomes and corresponding error location(s) for the parity check around a single site. Two or more errors are denoted generically as 2+.}\label{tab:2d1site}
\end{table}

\begin{table*}[]
\begin{tabular}{c|c|c||c|c|c||c|c|c||c|c|c||c|c|c|c||l}
$P^a_{(0,0)}$ & $P^b_{(0,0)}$ & $P^c_{(0,0)}$ & $P^a_{(0,1)}$ & $P^b_{(0,1)}$ & $P^c_{(0,1)}$ & $P^a_{(1,1)}$ & $P^b_{(1,1)}$ & $P^c_{(1,1)}$ & $P^a_{(1,0)}$ & $P^b_{(1,0)}$ & $P^c_{(1,0)}$ & $P^a_{(0,-1)}$& $P^a_{(-1,1)}$& $P^a_{(1,2)}$& $P^a_{(2,0)}$&error location  \\ \hline
+ & + & + & + & - & - & + & + & + & + & + & + & + &+&+&+&$s_{10}$\\ \hline
+ & + & + & + & - & - & + & - & - & + & - & - & + &+&+&+&$s_{11}$\\ \hline
+ & + & + & + & + & + & + & + & + & + & - & - & + &+&+&+&$s_{01}$\\ \hline
+ & - & - & + & - & - & + & + & + & + & - & - & + &+&+&+&$s_{00}$ \\ \hline
+ & + & + & + & - & - & + & + & + & + & + & + & +&+&+&-& $L^y_{20}$\\ \hline
+ & + & + & + & - & - & + & - & - & + & - & - & +&+&-&+& $L^x_{12}$\\ \hline
+ & + & + & + & + & + & + & + & + & + & - & - & +&-&+&+& $L^y_{01}$\\ \hline
+ & - & - & + & - & - & + & + & + & + & - & - & -&+&+&+& $L^x_{00}$ \\ \hline
\end{tabular}
\caption{Full parity check outcomes required to uniquely localize the errors on sites or external links opposite to the location of the $A_{\bf s}$ ancillas obtained using  four additional $P^a$ stabilizer measurement on neighboring sites. }\label{tab:2dparitycheck_nb}
\end{table*}

The extended 2D scheme presented here uses only standard stabilizer measurements over phase-flip encoded qubits and can thus be implemented in a fault tolerant way using similar techniques to those discussed in the 1D case. The main non-standard component of the scheme is the presence of the ancillae $A_{\bf s}$ which do not store the state of another qubit but the parity of a pair. The initial logical state can be initialized in a straightforward way by starting the system in a physical product state and making sure that any change performed to the state of the links connected to a given ancilla are also applied to the latter. For example, when implementing a hopping term which changes the flux on one of the links, the same change has to be enacted on the ancilla touching that link before continuing. In the $\bb{Z}_2$ case these extended operations can be achieved with just Clifford gates as they correspond simply to doubling the number of CNOT required.

The total qubit count for a square system with $4N_xN_y$ sites and $8N_xN_y$ links using a standard $[5,1,3]$ code, and neglecting ancillae required for the stabilizer measurements, would be $60N_xN_y$ while the scheme proposed in this work requires only $48N_xN_y$ qubits. Similarly to the 1D case discussed in previous sections, the proposed mapping has the additional benefit of allowing simpler implementations of transversal operations. The savings afforded by this scheme can be improved in a similar way as what we did for the one dimensional case: if one can tolerate a single fault in a larger region than the one depicted in panel (A) of Fig.~\ref{fig:2dlattice} then some of the ancilla registers $A_{\bf s}$ could be removed. For instance, if one extends the region where we can tolerate one error to enclose the four plaquettes neighboring a specific plaquette, a fault-tolerant scheme can be devised using only half of the $A_{\bf s}$ registers.

The present discussion of the 2D case can serve as a guideline to design similar error correction codes for higher dimensional geometries by adding additional ancilla registers to store the parity of subsets of links. Finally, extensions of the present scheme to $\bb{Z}_N$ or to U(1) with larger flux cutoffs could be constructed in principle using the same unary encoding map proposed for the one dimensional case.

\section{Discussion}
\label{sec:discussion}

Error correction involves extending the original Hilbert space of the system to a larger one and endowing the larger Hilbert space with local symmetries that define the codespace. Stabilizers of the codespace are in effect ``Gauss' Law" operators of that symmetry. 
We have outlined a simple fault tolerant algorithm which exploits local symmetries to reduce the space requirements for performing error correction on $\bb{Z}_2$ or U(1) LGT systems with a flux cutoff of $1$. The logical qubits were concatenated within a phase flip code to attain fault tolerance since phase flip errors commute with the Gauss' Law of these theories. 

As we have primarily investigated 1+1 dimensional systems with a flux cutoff of $1$, we outline the difficulties one encounters when attempting to generalize these procedures to arbitrary dimensions and flux cutoffs. 

We first note that for further levels of concatenation, our techniques incorporating Gauss's law can only be applied at the highest logical level where the bit-flip encoding is implemented instead of within every other level of the concatenation. Our approaches can conceivably be used to suppress simulation errors resulting from the approximation of the true time-evolution via methods such as Trotterization, but the impact of our approach on these errors is left for future work as not all them will lead to violations of Gauss's law. 

For arbitrary flux cutoffs and with non-dynamical matter fields on the sites, the classical CNOT gates in \autoref{fig:ECnondynamical} must be replaced with a classically controlled version of the $\hat{W}_s$ operation described in Sec.~\ref{subsec:gauss_construction}. These are controlled on the state of an appropriate site qubit and can be implemented with adder circuits such as those described in \cite{StrykerOracle_2019,ShawSchwinger_2020}. A fault-tolerant adder will however require non-Clifford operations for a binary integer encoding. In principle, these can be applied using standard techniques like state injection \cite{BravyiHaah} even with our logical encoding. A more scalable solution however is to instead use a unary integer encoding, for which a fault-tolerant incrementer can be implemented using only Clifford gates. The more efficient construction relying on ancilla qubits to compute the XOR between a site and a link variable shown in \autoref{fig:XOREquiv} can also be implemented in a straightforward way when using unary encoding. It has been shown recently that in the context of dynamical simulations, the cutoff only grows as $\text{polylog}(1/\varepsilon)$ in many physical situations~\cite{tong2021provably} for a fixed error $\varepsilon$. Thus the scaling in the number of qubits needed for integrating our error correction procedures with quantum simulation of such systems in a unary encoding is more optimal than the naive expectation of linear in the cutoff. However, existing unary encodings require non-stabilizer codes, so constructing generalized unary encodings which are stabilizer codes and exploring how the constructions given in \autoref{fig:GaussLawDynamical} generalize to this setting is currently under investigation. 


As discussed in Sec.~\ref{subsec:gauss_construction}, the construction proposed here for the 1D case using an alternating encoding with two qubits per even link and one qubit per odd link requires $9N$ qubits (excluding ancillas) for the theory with non-dynamical fermions and for pure gauge theory. This improvement in space complexity can also be extended to accommodate dynamical fermions by using Gauss' law to fix errors on the sites and requires $15N$ qubits, excluding ancillas. The 2D scheme presented in Sec.~\ref{sec:2dlgt} allows for an even more compact encoding where only $48N_xN_y$ qubits are required in the dynamical case and $36 N_x N_y$ for the pure gauge case for a lattice with $8N_xN_y$ links and $4N_x N_y$ sites. Such costs cannot be attained from error correcting individual qubits only. Moreover, while we considered only local Gauss's law checks, examining the links coming in and out of larger spatial regions can be used identify errors of Pauli weight more than 2. It would be interesting to investigate how the present constructions can be generalized to increase the distance of the effective bit-flip code that LGTs provide via Gauss's law. We also note that when multiple logical qubits are to encoded, an even lower physical-to-logical qubit ratio is achievable~\cite{Gottesman1996,Calderbank1997}.  The question of whether more efficient encodings could be found by following a more holistic approach as opposed to those using only Gauss' Law on local degrees of freedom like sites, links, and plaquettes is left for future work.

Most importantly, we do not claim to encode the full continuous gauge group, as this would violate the fact that finite dimensional quantum systems which correct erasure have no continuous symmetries but only a discrete subgroup thereof \cite{Faist2020}. As such, we only encode the U(1) symmetry group with a finite cutoff approximation. The construction of error correcting codes analogous to the ones presented here for other structure groups will therefore need to be tailored to their admissible discrete subgroups. 

We make the final observation that the loop string hadron formalism allows the diagonalization of Gauss' law for non-abelian gauge theories, converting them into analogues of those in abelian gauge theories \cite{Raychowdhury2020,Davoudi2021}. It is expected that this will be an important step in extending these algorithms to more complex theories where additional commuting constraints are needed. 

\begin{acknowledgements}

We thank Jesse Stryker and Martin Savage for useful discussions in the early stages of this work. This work was supported in part by the U.S. Department of Energy, Office of Science, Office of Nuclear Physics, Inqubator for Quantum Simulation (IQuS) under Award Number DOE (NP) Award DE-SC0020970 and by the DOE QuantISED program through the theory  consortium ``Intersections of QIS and Theoretical Particle Physics" at Fermilab. It was further supported by a grant from Google research award, and NW's theoretical work on this project was supported by the U.S. Department of Energy, Office of Science, National Quantum Information Science Research Centers, Co-Design Center for Quantum Advantage under contract number DE-SC0012704. Funded by the European Union under Horizon Europe 
Programme - Grant Agreement 101080086 — NeQST. Views 
and opinions expressed are however those of the 
author(s) only and do not necessarily reflect those 
of the European Union or European Climate, 
Infrastructure and Environment Executive Agency (CINEA). 
Neither the European Union nor the granting authority 
can be held responsible for them.

\end{acknowledgements}

\bibliography{oracleref}

\end{document}